\documentclass[12pt]{article}

\begin{document}

\title{Bayesian Conditioning, the Reflection Principle, and Quantum
  Decoherence\footnote{This article is dedicated to the memory of Itamar Pitowsky.  Already 10 years ago, he saw the potential in a Bayesian understanding of quantum probabilities and had a respect for our efforts like no one else \cite{Pitowsky1,Pitowsky2,Pitowsky3}.  The respect was mutual. For you, Itamar.}}

\author{Christopher A. Fuchs and R\"udiger Schack}

\date{\today}
\maketitle

\begin{abstract}
The probabilities a Bayesian agent assigns to a set of events typically change with time, for instance when the agent updates them in the light of new data. In this paper we address the question of how an agent's probabilities at different times are constrained by Dutch-book coherence. We review and attempt to clarify the argument that, although an agent is not forced by coherence to use the usual Bayesian conditioning rule to update his probabilities, coherence does require the agent's probabilities to satisfy van Fraassen's [1984] {\it reflection principle\/} (which entails a related constraint pointed out by Goldstein [1983]). We then exhibit the specialized assumption needed to recover Bayesian conditioning from an analogous reflection-style consideration.  Bringing the argument to the context of quantum measurement theory, we show that ``quantum decoherence'' can be understood in purely personalist terms---quantum decoherence (as supposed in a von Neumann chain) is not a physical process at all, but an application of the reflection principle.  From this point of view, the decoherence theory of Zeh, Zurek, and others as a story of quantum measurement has the plot turned exactly backward.
\end{abstract}

\section{Introduction}

At the center of most accounts of Bayesian probability theory \cite{Bernardo94} is the procedure of Bayesian
conditioning. By this we mean the following. Assume a Bayesian agent, at some
time $t=0$, has assigned probabilities $P_0(E)$, $P_0(D)$ and $P_0(E,D)$ to
events $E$ and $D$ and their conjunction. As long as $P_0(D)\ne0$, the
conditional probability of $E$ given $D$ is then
\begin{equation}   \label{eq:BayesianUpdating}
P_0(E|D) = \frac{P_0(E,D)}{P_0(D)} \;.
\end{equation}
Now assume that, at a later time $t=\tau$, the agent learns that $D$ is
true and updates his probability for $E$. We denote the agent's updated
probability by $P_\tau(E)$. Standard Bayesian conditioning consists of setting
\begin{equation}
P_\tau(E)=P_0(E|D)\;.
\label{NormaBayes}
\end{equation}

The rule Eq.~(\ref{NormaBayes}) can be viewed as a possible answer to the
general question of how an agent's probabilities at two different times should
be related. We will address this question from a personalist Bayesian
perspective
\cite{Bernardo94,Ramsey26,DeFinetti31,Savage54, DeFinetti90,Jeffrey04},
according to which probabilities express an agent's uncertainty, or degrees of
belief, about future events and acquire an operational meaning through
decision theory \cite{Bernardo94}. Although they are not determined by
agent-independent facts, personalist probability assignments are not
arbitrary. Dutch-book coherence \cite{Ramsey26,DeFinetti90,Jeffrey04} as a
normative principle requires an agent to try her best to make her numerical belief assignments conform to the
usual rules of the probability calculus. When coupled with the agent's overall
belief system, this is a powerful constraint \cite{Logue95}. Personalist
Bayesian probability is at the heart of Quantum Bayesianism, a radical new approach to the foundations of quantum mechanics developed by
Caves, Fuchs, Schack, Appleby, Barnum, and others. (See \cite{QBies1,QBies2} for an extensive reference list.)  The motivation for the
present investigation is to explore the relevance of Bayesian conditioning to the Quantum Bayesian program.

It was first pointed out by Hacking \cite{Hacking1967} that there is no
coherence argument that compels the agent to take into account the earlier
probabilities $P_0(E|D)$ when setting the later probabilities $P_\tau(E)$. Similar
points have been made by other authors (see, e.g.,~\cite{HowsonUrbach}).
Hacking was writing about the standard {\it synchronic\/} Dutch book
arguments, but the above statement remains true even for the {\it
  diachronic\/} Dutch book arguments, originally due to Lewis and first
reported by Teller \cite{Teller1973}. Without further assumptions, diachronic coherence does not
compel the agent, at $t=\tau$, to use the Bayesian conditioning rule
(\ref{eq:BayesianUpdating}).

The way diachronic coherence arguments connect
probability assignments at different times is more subtle. It is expressed
elegantly through van Fraassen's reflection principle \cite{vanFraassen1984},
which itself entails the related constraints of Shafer
\cite{Shafer1983} and Goldstein \cite{Goldstein1983}. The key idea behind the
reflection principle is to consider the agent's beliefs about his own future
probabilities, i.e., to consider expressions such as $P_0(P_\tau(E)=q)$. Shafer \cite{Shafer1983} put the point very nicely,
\begin{quotation}
\noindent This interpretation is based on the assumption that a person has subjective probabilities for how his information and probabilities may change over time.  This means we are concerned not with how the person {\em should\/} or {\em will\/} change his beliefs, but rather with what he believes about how these beliefs will change.  [Emphases added.]
\end{quotation}
The same idea underlies the approach this paper takes towards Bayesian conditioning and quantum decoherence.

In the next section, we present a detailed example where the agent appears
justified to depart from the Bayesian conditioning rule.  In
Section~\ref{sec:synchronic} we review the standard, synchronic, Dutch book
arguments and show why they do not imply Bayesian conditioning.
Section~\ref{sec:diachronic} introduces diachronic coherence and presents a
derivation of the reflection principle. In Section~\ref{sec:reflection} we show
that the Bayesian conditioning rule can be understood as a variant of the
reflection principle valid for a particular class of situations.
Section~\ref{sec:sirens} addresses an argument that has been advanced against
the reflection principle and shows that it is based on a misconception of the
role that coherence considerations have in probability assignments. Finally,
in Section~\ref{sec:decoherence} we give a natural application of the reflection
principle to decoherence in quantum mechanics from a Quantum Bayesian
perspective.

\section{Example: Polarization Data} \label{sec:polarization}

Consider a physicist running an experiment to discover the linear polarization of photons coming from a rather complicated optical device which he had built himself.  Perhaps he is convinced that every photon is produced the same way, only that he has forgotten which orientation $\theta$ he gave to a certain polarization filter deep within the set-up.  It might thus be easier to discover $\theta$ and recalibrate than to tear the whole thing apart and readjust.  A statistical analysis is in order.

Our experimenter will measure polarization for a sequence of $n$
individual photons and carry out a Bayesian analysis of the outcomes
consisting of a string, $s_n$, of zeros and ones.  The zeros stand for
horizontal polarization and ones for vertical polarization. Before starting, the experimenter records his probabilistic prior, which in the personalist
approach to probability adopted in this paper, represents his Bayesian degrees
of belief about the measurement outcomes. To be specific we assume that the
prior is {\em exchangeable}, i.e., of the form \cite{DeFinetti90,Caves02b}
\begin{equation}  \label{eq:deFinetti}
P_0(s_n) = \int_0^1 q_0(x) x^k (1-x)^{n-k} dx \;,
\end{equation}
where $k$ is the number of zeros in $s_n$, and $q_0(x)$ is a
probability density.  If the experimenter is completely ignorant of what orientation he had given the filter, he might assume $q_0(x)$ to be the constant density, but the precise form of $q_0(x)$ is of no great importance to
the argument below.

The prior $P_0(s_n)$  is a convex sum of binomial distributions $\{x,1-x\}$, with $x=\cos^2 \theta$. It is
symmetric in the sense that it is invariant under permutations of the
bits in $s_n$. By adopting this prior, the experimenter judges that
the order in which the photons arrive is irrelevant to his analysis.  To a Bayesian, this is in fact the operational meaning that all photons are ``produced the same way.''  A simple consequence of this is that a posterior probability calculated from this prior by Bayesian conditionalization after some number of trials will not depend on the order of the zeros and ones found in those trials.

%For instance, let $E$ be
%the event that the $n$th photon has polarization 0, let $s_{n-1}$ denote the
%string of outcomes for the first $n-1$ photons, and let $k$ be the number of
%zeros in $s_{n-1}$. The joint (prior) probability for $s_{n-1}$ and $E$ is
%\begin{equation}
%P_0(s_{n-1},E) = \int_0^1 q_0(x) x^{k+1} (1-x)^{n-k-1} dx \;,
%\end{equation}
%and the conditional probability of $E$ given $s_{n-1}$ is
%\begin{equation}  \label{eq:priorconditional}
%P_0(E|s_{n-1}) = \frac{P_0(s_{n-1},E)}{P_0(s_{n-1})} \;,
%\end{equation}
%where $P_0(s_{n-1})$ is obtained by substituting $n-1$ for $n$ in
%Eq.~(\ref{eq:deFinetti}). The conditional
%probability~(\ref{eq:priorconditional}) will depend only on $n$ and
%$k$, not on any other details of the string $s_{n-1}$.

Suppose now that the experimenter observes 4000 trials, and finds very nicely that the number of zeros and ones
is not very far from 2000 each.  %The experimenter says to himself, ``Great!  The polarization filter was oriented a simple 90 degrees to what I had intended.'' 
Technically this means that if the experimenter updates his prior to a posterior by Bayesian conditionalization, the posterior for the next $n$ bits will be
\begin{equation}
P_{\rm posterior}(s_n) = \int_0^1 q'(x) x^k (1-x)^{n-k} dx \;,
\end{equation}
where $q'(x)$ is a function on the interval $[0,1]$ peaked very near 1/2. If the experimenter were asked to bet
on the next bit, this probability distribution would advise him to bet at close to even odds.

%Now assume that $n=1001$, and the experimenter has observed $k=500$, i.e., the
%string $s_{n-1}$ contains the same number of zeros and ones. This implies that
%\begin{equation}
%P_0(E|s_{n-1})=0.5 \;.
%\end{equation}
But now consider the following fantastic scenario. Suppose the
experimenter becomes aware that the string $s_{n}$ he accumulated is identical to the first
4000 bits of the binary expansion of $\pi$! Any sane person would be flabbergasted.  Even though the experimenter built the device with his own hands, he would surely wonder what was up.  Perhaps one of his lab partners has played an immense joke on him?

The following question becomes immediate: If the experimenter is rational, how should he bet {\it now\/} on the next bit?  Sticking doggedly with Bayesian conditioning, he would be advised to use near even odds, just as before.  But the number $\pi$ is too significant to ignore:  His heart says to bet with
\begin{equation}   \label{eq:conditioning}
P_\tau(E) = 0.99
\end{equation}
on the event $E$ that the next bit equals the 4001th bit
of the binary expansion of $\pi$. 
Our experimenter faces a stark choice: He can either ignore his heartfelt belief and use the value
$P_\tau(E)\approx 1/2$ obtained from the conditioning rule, or he can ignore the conditioning rule and use the
value $P_\tau(E)=0.99$ representing what he really feels. Both choices are deeply
problematical: the second one seems to be incoherent because it is
contradicted by the usual understanding of the formalism, whereas the first one seems to be ignoring common
sense to the point of being foolish.

Is the second choice really incoherent however?  Does it violate a reasonable normative
requirement? We will see in the next sections that this is not the
case.

\section{Synchronic Dutch Book} \label{sec:synchronic}

A simple way to give an operational meaning to personalist Bayesian
probabilities is through the agent's betting preferences. %\footnote{To quote Pitowsky \cite{Pitowsky1}, ``This simple scheme suffers from various weaknesses, and better ways to associate epistemic probabilities with gambling have been developed ...  For a more sophisticated way to associate probability and utility see Savage \cite{Savage54}.''}
When an agent assigns probability $p$ to an event $E$, she regards $\$p$ to be the fair
price of a standard lottery ticket that pays $\$1$ if $E$ is true. In other
words, an agent who assigns probability $p$ to the event $E$ regards both
buying and selling a standard lottery ticket for $\$p$ as fair transactions;
for her, the ticket is worth $\$p$.

In the following, we will call a set of probability assignments {\it
  incoherent\/} if it can lead to a sure loss in the following sense: there
exists a combination of transactions consisting of buying and/or selling a
finite number of lottery tickets which (i) lead to a sure loss and (ii) the
agent regards as fair according to these probability assignments. A set of
probability assignments is {\it coherent\/} if it is not incoherent. We accept
as a normative principle that an agent should aim to avoid incoherent
probability assignments.

The standard, synchronic, Dutch book argument \cite{Ramsey26,DeFinetti90,Jeffrey04}
shows that an agent's probability assignments $P_0$ at a given time are
coherent if and only if they obey the usual probability rules, i.e., $0\le
P_0(E)\le1$ for any event $E$; $P_0(S)=1$ if the agent believes the event $S$
to be true; and $P_0(E\vee D)=P_0(E)+P_0(D)$ for any two events $E$ and $D$
that the agent believes to be mutually exclusive.

In this approach, conditional probability is introduced as the fair price
of a lottery ticket that is refunded if the condition turns out to be false.
Formally, let $D$ and $E$ be events, and let $\$q$ be the price of a
lottery ticket that pays $\$1$ if both $D$ and $E$ are true, and $\$q$ (thus
refunding the original price) if $D$ is false. For the
agent to make the conditional probability assignment $P_0(E|D)=q$ means that she
regards $\$q$ to be the fair price of this ticket.

It is then a consequence of Dutch-book coherence that the product rule
$P_0(E,D)=P_0(E|D)P_0(D)$ must hold \cite{Jeffrey04}. In other words, conditional probability
assignments violating this rule are incoherent. If $P_0(D)\ne0$, we obtain
 Bayes's rule,
\begin{equation} \label{eq:bayesRule}
P_0(E|D)=\frac{P_0(E,D)}{P_0(D)} \;.
\end{equation}
It is worth pointing out that Bayes's rule emerges here as a theorem, combining
terms that are defined independently, in contrast to the common axiomatic
approach to probability theory where Eq.~(\ref{eq:bayesRule}) is used as the
definition of conditional probability.

The above shows that a coherent agent must use Bayes's rule to set the
conditional probability $P_0(E|D)$. The value of $P_0(E|D)$ expresses what
ticket prices the agent regards as fair at time $t=0$, i.e., before she finds
out the truth value of either $D$ or $E$. It says nothing about what ticket
prices she will regard as fair at some later time $t>0$.

In particular, assume that, at some time $t=\tau>0$, the agent learns that $D$
is true and updates her probabilities accordingly. Denote by $P_\tau(E)$ the
agent's updated probability of $E$, meaning that she now regards $\$P_\tau(E)$
as the new fair price of a ticket that pays $\$1$ if and only if $E$ is true.
Nothing in the Dutch book argument sketched above implies that $P_\tau(E)$
should be equal to $P_0(E|D)$ \cite{Hacking1967}.  All probabilities used in the
argument are the agent's probabilities at time $t=0$; they are defined via
ticket prices for bets on $E$, $D$ and their conjunction which the agent regards as
fair at $t=0$.  The Dutch book argument leading to Eq.~(\ref{eq:bayesRule}) is
a {\em synchronic\/} argument. It does not connect in any way the agent's
probability assignments at $t=0$ and $t=\tau$. In particular, it does not imply
that the agent has to use Bayesian conditioning to update her probabilities.
In the next section we will see what connection between the agent's
probability assignments at different times actually is implied by
diachronic Dutch book arguments.

\section{Diachronic Dutch Book} \label{sec:diachronic}

To set the scene, we consider an investor who today buys 500 shares of some
company at a price of $\$20$ each, which he regards as a fair deal. The next
day, his appreciation of the market has changed, and he sells his 500 shares
at $\$18$ each, which now, given the new situation, he again regards as a fair
deal. Despite the fact that the investor makes a net loss of $\$1000$, he
does not behave irrationally. By selling his shares at a lower price
on the next day, he simply cuts his losses.

But what if the investor is certain today that tomorrow he will regard $\$18$
as the fair price for a share? It would then be foolish
for him to buy, today, 500 shares for $\$20$ each, because he is
certain that tomorrow he would be willing to sell the shares for $\$18$
each, leading to a net loss of $\$1000$. As a matter of fact, buying shares at
any price above $\$18$ today would be foolish in this situation, as would be
selling shares today at any price below $\$18$.

In the above example, we have assumed that money has the same utility for the
investor today and tomorrow, i.e., we have assumed a zero interest rate.  This
is an assumption we will make throughout the present paper. More precisely, we
will assume that the time at which she receives a sum of money is irrelevant to
a Bayesian agent.

In probability language, what we have just described is the following. Assume
$P_0(E)$ is an agent's probability at $t=0$ of some event $E$. The agent buys
a lottery ticket that pays $\$1$ if $E$ is true, for $\$P_0(E)$ which she
regards as the fair price. At a later time $t=\tau$, she updates her beliefs.
Her probability for $E$ is now $P_\tau(E)$, which happens to be less than
$P_0(E)$. At this point, the agent decides to cut her losses by selling the
ticket for $\$P_\tau(E)$. Despite the net loss, there is nothing irrational
about the agent's transactions.

But now suppose that at $t=0$ the agent is certain that, at $t=\tau$, her
probability of $E$ will be $q$, where $q\ne P_0(E)$. In the case $q<P_0(E)$,
this means that, at $t=0$, she is willing to buy a ticket for $\$P_0(E)$
although she already knows that later she will be willing to sell it for the
lower price $\$q$. In the case $q>P_0(E)$, it means that, at $t=0$, she is
willing to sell a ticket for $\$P_0(E)$ although she already knows that later
she will be willing to buy the same ticket for the higher price $\$q$. In both
cases, already at $t=0$ the agent is certain of a sure loss.

This simple scenario contains the main idea of van Fraassen's diachronic Dutch book
argument. Similar to the synchronic case discussed in the previous section, we
call an agent's probability assignments incoherent if there exists a
combination of transactions consisting of buying and/or selling a finite
number of lottery tickets at two different times such that (i) {\em already at the
earlier time}, the agent is sure of a net loss; and (ii) each transaction is
regarded as fair by the agent according to her probability assignments {\em at the
time\/} the transaction takes place. We will continue to accept as a normative
principle that an agent should aim to avoid incoherent probability
assignments.

To turn our simple scenario into the full-fledged diachronic Dutch-book
argument, we only need to relax the assumption that at $t=0$ the agent is
{\it certain\/} that $P_\tau(E)=q$. Instead, we assume that
\begin{equation}  \label{eq:beliefAboutBelief}
P_0\Big(P_\tau(E)=q\Big) > 0 \;,
\end{equation}
i.e, at $t=0$ the agent believes with some positive probability that at
$t=\tau$ her probability of $E$ will be equal to $q$. We will now show
that this implies the agent's probability assignments are incoherent
unless
\begin{equation}  \label{eq:reflection}
 P_0\Big(E\,\Big|\,P_\tau(E)=q\Big) = q \;,
\end{equation}
i.e., unless at $t=0$ the agent's conditional probability of $E$, given that
$P_\tau(E)=q$, equals $q$. This is van Fraassen's {\it reflection principle\/}
\cite{vanFraassen1984}.

To derive the reflection principle, denote by $Q$ the proposition
$P_\tau(E)=q$, i.e., the assertion that at time $t=\tau$, the agent will regard
$\$q$ as the fair price for a ticket that pays $\$1$ if and only if $E$ is
true. The inequality (\ref{eq:beliefAboutBelief}) thus becomes $P_0(Q)>0$.  To
establish that coherence implies the reflection principle
(\ref{eq:reflection}), one must show that the assumption $P_0(E|Q)\ne q$ leads to
a sure loss for an appropriately chosen set of bets no matter what outcomes occur for the events considered.

As a warm-up to gain intuition, suppose that $P_0(E|Q)>q$ and that $Q$ is true. This means that at $t=0$, the agent is willing to
buy a ticket for $\$P_0(E|Q)$ that pays $\$1$ if both $Q$ and $E$ are true, and refunds the ticket price if $Q$ is false. But, because of $Q$'s truth, this ticket will further be equivalent to a ticket that pays $\$1$ if $E$ is true.  Finally, the truth of $Q$ also implies that at $t=\tau$ the agent will be willing to sell this ticket for $\$q$, which is less than what she paid for it. In other words, if $Q$ is true, the agent is sure to lose $\$d$, where $d=P_0(E|Q)-q$.

But this simple argument---illustrative though it may be---is not a full-fledged proof of incoherence.  To get a full proof we need to show that the agent is sure of a loss not only when $Q$ turns out to be true, but also when $Q$ turns out to be false. For this it is sufficient to consider an alternate scenario where there is a side bet on $Q$, such that the agent loses some amount if $Q$ is false, and wins
less than $\$d$ if $Q$ is true. Such a side bet may be realized by a
lottery ticket that pays $\$d/2$ if $Q$ is true, which the
agent is willing to buy for $\$P_0(Q)d/2$.

We have thus the following combination of transactions, each of which the
agent regards as fair at the time it takes place:
\begin{itemize}
\item[(i)] to buy, at $t=0$, for \$$P_0(E|Q)$, a ticket that pays $\$1$ if
  both $Q$ and $E$ are true, and refunds the ticket price if $Q$ is false;
\item[(ii)] to buy, at $t=0$, for $\$P_0(Q)d/2$, a lottery ticket that pays
  \$$d/2$ if $Q$ is true;
\item[(iii)] if $Q$ is true (i.e., if $P_\tau(E)=q$), to sell, at $t=\tau$, for
  \$$q$, a lottery ticket that pays \$1 if $E$ is true.
\end{itemize}
Already at $t=0$, the agent knows that these transactions result in a net
loss, equal to $\$(P_0(Q)+1)d/2$ if $Q$ is true, and $\$dP_0(Q)/2$ if $Q$ is
false. We have thus shown that the assumption $P_0(E|Q)>q$ implies that the
agent's probability assignments are incoherent.

The final piece of a proof is to consider the case $P_0(E|Q)<q$.  By reversing the signs of all
transactions above, it is easy to see that this case leads to a sure loss in exactly the same way.
Putting these two cases together, this completes the full derivation of the
reflection principle.

%% If my probabilities today do not obey the reflection principle, using these
%% probabilities to decide if a set of transactions is fair can thus lead the
%% agent to a sure loss. Coherence thus implies the reflection principle.

The coherence condition of Shafer \cite{Shafer1983} and Goldstein \cite{Goldstein1983} follows by a simple application of synchronic coherence along with the reflection principle.  Suppose the agent instead of contemplating a single $Q = [P_\tau(E)=q] $ for what she will believe of $E$ at $t=\tau$, contemplates a range of mutually exclusive and exhaustive propositions $\{Q\}$ to which she assigns probabilities $P_0(Q)$.  Then, straightforward synchronic coherence requires
\begin{equation}
P_0(E)=\sum_Q P_0(Q) P_0(E|Q)
 = \sum_q P_0\Big(P_\tau(E)=q\Big)\,P_0\Big(E\,|\,P_\tau(E)=q\Big)   \;,
\end{equation}
for which reflection in turn implies
\begin{equation}
P_0(E) = \sum_q P_0\Big(P_\tau(E)=q\Big)\, q\;.
\label{Hermeneutic}
\end{equation}
This implication of the reflection principle will turn out to be particularly important for our exposition of quantum decoherence.

\section{Bayesian Conditioning in Reflectional Terms}  \label{sec:reflection}

% It is widely recognized that the validity of the Bayesian conditioning rule
% depends on further assumptions (see, e.g., \cite{HowsonUrbach}).

% 6. Conditions for updating: I am certain now that if D occurs my probability
% of $E$ will be P. Then P must be $P(E|D)$ (or so).

% Another idea: $P_0(P_\tau=q | D))=1$ for some $\tau$. This is the condition
% under which the Bayesian rule holds. Period. This is not saying that if $D$,
% $P_\tau$ will definitely be $q$. It simply expresses the agents degree of
% belief about the matter at $t=0$. Anything else: category mistake.

% UPDATING STRATEGY: SKYRMS ET AL ARE LESS COMPELLING THAN THE DRAFT
% SUGGESTS. HAVING AN UPDATING STRATEGY DOES NOT IMPLY CERTAINTY THAT THE
% ACTUAL BELIEF AT T=TAU IS THE ONE GIVEN BY THE UPDATING STRATEGY.

The reflection principle is a constraint on an agent's present beliefs about her future probability assignments.  It does not directly provide an explicit rule for assigning probabilities, either for the present ones or the future ones. An agent whose
probabilities violate the reflection principle is incoherent and
should strive to remove this incoherence. The reflection principle
does not provide a recipe for how to do this.

One way in which the agent can achieve coherence is by adopting an ``updating
strategy'' \cite{Skyrms1987} based on the Bayesian conditioning rule. We will now
explore to what extent the Bayesian conditioning rule follows for such a strategy in a way analogous to the
reflection principle---that is, in a way ``concerned not with how the person should or will change his beliefs, but rather with what he believes about how these beliefs will change'' \cite{Shafer1983}.

Let $E$ and $D$ be events, and let $P_0(E)$, $P_0(D)$ and $P_0(E|D)$ denote
the agent's respective probabilities at $t=0$.  Assume that the truth value of
$D$ will be revealed to the agent at $t=\tau$. Suppose she now adopts the strategy
that, if at $t=\tau$ she learns that $D$ is true, her updated probability of
$E$, denoted by $P_\tau(E)$, will be given by some value $q$, $0\le q\le1$.

The above can be phrased in terms of the agent's probabilities at $t=0$. For
her to adopt this strategy simply means that
she is certain that, if $D$ turns out to be true, she will make the
probability assignment $P_\tau(E)=q$, i.e.,
\begin{equation}  \label{eq:conditionalCertainty}
P_0\Big(P_\tau(E)=q\Big|D\Big) = 1\;.
\end{equation}
This statement about the agent's current belief about her future probability
captures the essence of Bayesian conditioning. Together with diachronic
coherence it implies that
\begin{equation}  \label{eq:reflectionVariant}
q = P_0(E|D)\;,
\end{equation}
i.e., the Bayesian conditioning rule. Presented in this way, it can be
regarded as a variant of the reflection principle, valid whenever the
condition (\ref{eq:conditionalCertainty}) holds.

To derive Eq.~(\ref{eq:reflectionVariant}), we consider again the combinations
of bets introduced in Sec.~\ref{sec:diachronic}, but with the event $D$
replacing $Q$ throughout. We assume first that $P_0(E|D)>q$ and define
$d=P_0(E|D)-q$. The transactions are
\begin{itemize}
\item[(i)] to buy, at $t=0$, for \$$P_0(E|D)$, a ticket that pays $\$1$ if
  both $D$ and $E$ are true, and refunds the ticket price if $D$ is false;
\item[(ii)] to buy, at $t=0$, for $\$P_0(D)d/2$, a lottery ticket that pays
  \$$d/2$ if $D$ is true;
\item[(iii)] if $D$ is true, to sell, at $t=\tau$, for
  \$$q$, a lottery ticket that pays \$1 if $E$ is true.
\end{itemize}
At $t=0$, the agent is certain that these transactions result in a net
loss, equal to $\$(P_0(D)+1)d/2$ if $D$ is true, and $\$P_0(D)d/2$ if $D$ is
false. At $t=0$, the agent regards (i) and (ii) as fair transactions, and
because of Eq.~(\ref{eq:conditionalCertainty}), she is certain that at
$t=\tau$ she will regard (iii) as a fair transaction. We have thus shown that
the agent's probabilities are incoherent if $P_0(E|D)>q$. The case
$P_0(E|D)<q$ is similar. Thus coherence implies that $P_0(E|D)=q$, as
required.

The key assumption in this derivation, expressed by
Eq.~(\ref{eq:conditionalCertainty}), is that the agent can identify an event
$D$ that she expects to determine her future beliefs. There are more general
updating strategies that are not of this form. Jeffrey's probability
kinematics \cite{Jeffrey1965} is such an example. Probability kinematics is a
coherent updating strategy \cite{Skyrms1987} which does not make use of the
Bayesian conditioning rule, but it too can be put in reflectional terms as we did with Bayesian conditioning.

Actually, one can go still further along these reflectional lines if one strengthens the assumption in Eq.~(\ref{eq:conditionalCertainty}) to also make a direct identification between the possible values for $P_\tau(E)$ and the $D$, i.e., that there is bijection between them.  In such a case, one can say that Bayesian conditioning {\em follows directly\/} from the reflection principle.  For then,
\begin{equation}
P_0(E|D)=P_0(E|D,Q)=P_0(E|Q)
\end{equation}
by standard synchronic logic, and $P_0(E|Q)=q$ by reflection.

%Bayesian conditioning can
%thus be regarded as an instance of the reflection principle, valid for a
%limited class of updating strategies.

The discussion above is entirely in terms of the agent's beliefs at $t=0$.
What if, at $t=\tau$, after learning that $D$ is true, the agent
re-analyses the situation, possibly taking into account circumstances she was
not aware of at $t=0$, and concludes that her new probability, $P_\tau(E)$,
differs from $P_0(E|D)$. Does this imply that her probability assignment is
incoherent? The answer is no. Coherence is a condition about an agent's {\it
  current beliefs}, including her beliefs about her future probability
assignments. In the above scenario, the agent's beliefs at $t=\tau$ are
coherent as long as $0\le P_\tau(E)\le1$. Nothing in the Dutch book argument
implies that the agent's actual probabilities at $t=\tau$ are constrained by
her probabilities at $t=0$, which supports the conclusion of
Sec.~\ref{sec:polarization} that there is no conflict with coherence for an
experimenter who assigns $P_\tau(E)=0.99$ although the Bayesian conditioning
rule appears to mandate $P_\tau(E)\approx 1/2$.

\section{Sirens, Car Keys, and Married Couples} \label{sec:sirens}

We have seen in the previous section that one way of satisfying the reflection
principle and thereby avoiding incoherence is to set your future probabilities
in terms of your current probabilities via Bayesian conditioning. The form of
the reflection principle, however, suggests a different way of
proceeding. Since Eq.~(\ref{eq:reflection}) expresses a constraint on a
current probability, conditioned on a future probability assignment, one could
take the future probability as given and set the current probability in terms
of it, thus reversing the usual direction of Bayesian updating. This can be a
useful and legitimate procedure. An important application will be given in
Sec.~\ref{sec:decoherence} below.

If the reflection principle is taken as a rule to set future probabilities in
terms of current probabilities, it can lead to decisions that appear
irrational \cite{Maher1992,Christensen1991,Skyrms1993,GreenHitchcock1994}. A classic example
\cite{vanFraassen1995} is provided by the story of Ulysses and the Sirens. Ulysses
knows that tomorrow, as soon as he is within earshot of the Sirens, he will
make a catastrophic decision.  Does the reflection principle force him to
endorse this catastrophic decision today?

Since analysing this story would involve a discussion of utility, here is
another famous example \cite{Christensen1991}. I know that I will get drunk this
evening and that I will assign probability $10^{-6}$ to the event $E$ of my
causing an accident while driving home late at night. Does reflection imply
that I must assign probability $10^{-6}$ to the event $E$ now?

Examples like this have led, e.g., Christensen \cite{Christensen1991} to the conclusion
that the reflection principle is unsound. This conclusion stems from a
confusion about the role of coherence arguments, however. The reflection principle can
be regarded as a tool to detect incoherence. The Dutch book arguments show
that incoherent probability assignments have the potential to lead to
catastrophic consequences. This justifies accepting a normative rule that
an agent should adhere to the reflection principle in order to avoid
incoherence. The reflection principle does not, however, give a prescription
for setting probabilities, either today's in terms of tomorrow's or vice
versa.  There is a range of options for the agent once she has detected an
incoherence, as we will now illustrate.

Suppose that, in the example above, my initial conditional probability
for an accident if I drive home drunk is 0.01. Suppose further that I
am certain that I will get drunk, and that my probability for an
accident will then be $10^{-6}$.  These probabilities violate the
reflection principle. My probability assignments are therefore
incoherent. One way of avoiding this incoherence would be to decide
not to get drunk, which would mean assigning probability 0 to this
event and therefore restore coherence. There is another very practical
solution, which is to give my car keys to a trusted friend before I
start drinking. My probability assignments will still be incoherent,
but I will be unable to act on them.

Ulysses's solution, 3000 years ago, was very similar. He ordered his men to
chain him to the mast of his ship. His men were to plug their ears. He
accepted incoherence, but prevented himself from acting on his incoherent
probability assignments. His men achieved coherence by reducing the
probability of hearing the Sirens to zero. Coherence is an ideal one should
always strive for. Incoherent probability assignments have the potential to
lead to catastrophic consequences. If one can't achieve coherence, one should
give up the car keys, plug one's ears or chain oneself to a mast.

In his article contra reflection, Christensen \cite{Christensen1991} pointed out
that the reflection principle is very similar to a related principle which he
called {\it solidarity}. Consider husband and wife who share a bank account.
Denote the husband's probabilities by $P_h$ and the wife's by $P_w$, and
consider some event, $E$.  Solidarity is the principle that, given that
the wife's probability of $E$ is $q$, the husband's probability must also be
$q$, i.e.,
\begin{equation}
P_h\Big(E \Big| P_w(E)=q \Big) = q \;.
\end{equation}
Violating solidarity leads to a sure loss for the joint bank account exactly
like in the diachronic Dutch book argument for the reflection principle.

Christensen argued that the solidarity principle is clearly
absurd. This may be a case of confusion between normative and
descriptive rules.  The solidarity principle is a normative principle
and does not claim that actual agents' probability assignments are
always compatible with it. What it says instead is that, to avoid
potential catastrophic consequences for their common bank account,
husband and wife must strive for coherence. The solidarity principle,
or more generally, the reflection principle, provides a tool to detect
incoherence.  It is then up to the agents how to resolve the
incoherence.  The husband might give, the wife might give, or they might compromise after debating all the relevant issues.  The key point is that deliberation is to their mutual benefit, and coherence is their goal.

\section{Quantum Decoherence} \label{sec:decoherence}

%Up to now we have looked at two ways of using the reflection principle. In
%Sec.~\ref{sec:reflection}, we showed how the reflection principle leads to
%strategies for updating tomorrow's probabilities by conditionalization. 
In the last section, we described how the reflection principle can be
used to detect incoherence and thus to avoid catastrophic consequences. In
this section, we will see that there is a generic situation in quantum theory where the
reflection principle is used directly to set today's probabilities in terms of tomorrow's.

We will look at a standard quantum measurement situation \cite{NielsenChuang}
from the perspective of Quantum Bayesianism, according to
which all quantum states, pure or mixed, represent an agent's degrees of
belief about future measurement outcomes. Assume an agent has, at time $t=0$,
assigned a quantum state (i.e., density operator) $\rho_0$ to a quantum system. She
intends to perform two measurements on the system, the first one at time
$t=\tau>0$, the second one at a still later time $t=\tau^\prime$. She describes the first
measurement by a collection of trace-decreasing completely positive maps
$\{{\cal F}_i\}$, each corresponding to a potential outcome, $i$, for the
first measurement. These completely positive maps determine the agent's probabilities $P_0(i)={\rm tr}[{\cal F}_i(\rho_0)]$, at time $t=0$, for the outcomes $i$, but they also determine the states she will assign to the system after the measurement:  If outcome $i$, then $\rho_\tau=P_0(i)^{-1}{\cal F}_i(\rho_0)$.

To describe the second measurement, it is enough to use a POVM, i.e., positive operator valued measure, $\{E_j\}$, since we will not be considering any further measurements after it. In this description each positive operator $E_j$ corresponds to a potential outcome, $j$, for the second measurement. If $\rho_\tau$ is the agent's system state at time $t=\tau$, then her
probabilities, at $t=\tau$, for the outcomes $j$ are given by $P_\tau(j)={\rm tr}(E_j\rho_\tau)$.

%When the agent describes the first measurement by the collection of maps
%${\cal F}_i$, she effectively commits to an updating strategy consisting of
%setting $\rho_\tau=P_0(i)^{-1}{\cal F}_i(\rho_0)$ if the measurement outcome is
%$i$. Using the formulation introduced in Sec.~\ref{sec:reflection}, this
%updating strategy can be expressed as
%\begin{equation}  \label{eq:quantumCertaintyRho}
%P_0\Big(\rho_\tau=P_0(i)^{-1}{\cal F}_i(\rho_0)\; \Big|\; i \Big)=1
%\end{equation}
%or, directly in terms of the outcomes $j$ of the second measurement,
%\begin{equation}  \label{eq:quantumCertaintyP}
%P_0\Big(P_\tau(j)={\rm tr}[E_jP_0(i)^{-1}{\cal F}_i(\rho_0)] \;\Big| \;i \Big)=1
%\;.
%\end{equation}
%Equation~(\ref{eq:quantumCertaintyP}) has exactly the form of
%Eq.~(\ref{eq:conditionalCertainty}). The variant of the reflection principle
%discussed in Sec.~\ref{sec:reflection} thus implies that the agent's
%probabilities are incoherent unless
%\begin{equation}
%{\rm tr}[E_jP_0(i)^{-1}{\cal F}_i(\rho_0)] =P_0(j\,|\,i) \;.
%\end{equation}

Now suppose our agent is confronted at time $t=0$ with a bet concerning the outcome $j$ at $t=\tau^\prime$.  How should she gamble without having yet performed the measurement at $t=\tau$?  We can read the answer straight off the reflection principle as written in the form of Goldstein and Shafer, Eq.~(\ref{Hermeneutic})---remember here that $P_\tau(j)$ is implicitly dependent upon $i$:
\begin{equation}
P_0(j) = \sum_i  P_0(i) P_\tau(j) =
\sum_i {\rm tr} [E_j{\cal F}_i(\rho_0)] \;.
\end{equation}
Cleaning this up a bit, we can write:
\begin{equation}
P_0(j) = {\rm tr} [E_j\sum_i{\cal F}_i(\rho_0)] = {\rm tr} (E_j\rho'_0) \;,
\end{equation}
where
\begin{equation}   \label{eq:decoherence}
\rho'_0 = \sum_i{\cal F}_i(\rho_0) \;.
\end{equation}
What we have shown here is that the reflection principle entails that the
agent can obtain her probabilities at $t=0$ for the outcomes of the second
measurement from the density operator $\rho'_0$. The state $\rho'_0$, which
has the form of a ``decohered'' state, {\it is\/} the agent's
quantum state at $t=0$ {\it as far as the second measurement is concerned}.

These conclusions are valid for any pair of measurements, but a little more
can be said if the POVM $\{E_j\}$ is informationally complete, i.e., if the
state $\rho_\tau$ is fully determined by the probabilities $P_\tau(j)$. In
this case 
%Eq.~(\ref{eq:quantumCertaintyRho}) is implied by Eq.~(\ref{eq:quantumCertaintyP}), and 
$\rho'_0$ as defined in Eq.~(\ref{eq:decoherence}) is the only
density operator that gives rise to the probabilities $P_0(j)$ required by
the reflection principle.

Equation (\ref{eq:decoherence}) takes a perhaps more familiar form if the
first measurement is a von Neumann measurement and the updating is given by
the L\"uders rule. In this case the action of the maps ${\cal F}_i$ on the state
$\rho_0$ can be written as ${\cal F}_i(\rho_0) = \Pi_i\rho_0\Pi_i$, where the
$\Pi_i$ are projection operators, and Eq.~(\ref{eq:decoherence}) becomes
\begin{equation}  \label{eq:Lueders}
\rho'_0 = \sum_i \Pi_i\rho_0\Pi_i \;.
\end{equation}

A common attitude about quantum measurement is that it is something that demands a detailed physical explanation.  Much of the folklore since the publication of John von Neumann's 1932 book {\sl Mathematical Foundations of Quantum Mechanics\/} is that a quantum measurement is something that occurs in two steps: First, there is a kind of ``pre-measurement'' where the quantum system becomes entangled with a measuring device.  Secondly, there is a ``selection'' of one of the entangled state's components; this is what singles out a particular measurement result.

The trouble with this description, however, is that the entangled wave function, with its freedom to be expressed in any
bipartite basis, does not have enough structure to specify how it should be decomposed so that such a ``selection'' can
be effected. The theory of quantum decoherence, developed by Zeh,
Zurek, and others \cite{Schlosshauer07}, attempts to overcome this deficiency in the
von Neumann story by supplementing it with a further story of interaction
between the measuring device and an environment: The idea is that the specific form of
the interaction with the environment specifies how the joint state of system
plus device ought to be decomposed. In this picture, the decoherence process preceding the ``selection'' step leads to
a state of the form~(\ref{eq:Lueders}), or more generally,
(\ref{eq:decoherence}).  What remains mysterious in this picture, however, is
``the selection step'' itself.  Decoherence theorists usually leave that question
aside, implicitly endorsing one variety or another of an Everettian
interpretation of quantum mechanics.  

In contrast, the Quantum Bayesian view of quantum theory leaves most of the usual von Neumann story aside: Instead of taking quantum states and unitary evolution as the ontic elements to which the theory refers, it takes the idea of an individual agent's decisions and experience as the theory's real subject matter. In this view, the process called ``quantum measurement'' is nothing other than an agent acting upon the world and experiencing the consequences of her actions. For a Quantum Bayesian, the only {\em physical\/} process in a quantum measurement
is what was previously seen as ``the selection step''---i.e., the agent's action on the external world and its
unpredictable consequence for her, the data that leads to a new state of
belief about the system.

Thus, it would seem there is no foundational place for decoherence in the
Quantum Bayesian program. And this is true. Nonetheless, in the two-time
measurement scenario we described above, there is a coherent state assignment at time $t=0$ for the second measurement that mimics a belief in decoherence.
This is simply a consequence of the implications of the reflection
principle. The ``decohered'' state $\rho'_0$ is not the agent's state after
she has made the first measurement (that would have been {\it one\/} of the $\rho_\tau$ depending upon the $i$ found). It is not the state resulting
from the measurement interaction before the selection step takes place as the
decoherence program would have it (nothing is so intricately modeled here). It is simply a quantum state the agent uses
at time $t=0$ before the first measurement to make decisions regarding the outcomes of the
second measurement.

That is the story of decoherence from the Quantum Bayesian perspective.
Decoherence does not come conceptually before a ``selection,'' but
rather is predicated on a time $t=0$ belief regarding the possibilities for the next quantum state at time $t=\tau$.
Decoherence comes conceptually {\em after\/} the recognition of the future
possibilities. In this sense the
decoherence program of Zeh and Zurek \cite{Schlosshauer07}, regarded as an attempt to contribute to our understanding of quantum measurement, has the story exactly backward.

\section{Acknowledgements}

We thank Lucien Hardy for persisting that the example in Section 2 {\it should be\/} important to us.  We thank Matthew Leifer for bringing the work of Goldstein \cite{Goldstein1983} to our attention, which derivatively (and slowly) led us to an appreciation of van Fraassen's reflection principle \cite{vanFraassen1984}; if we were quicker thinkers, this paper could have been written six years ago.  This work was supported in part by the U.~S. Office of Naval Research (Grant No.\ N00014-09-1-0247).

\end{document}